\documentclass[conference,10pt,twoside,twocolumn]{IEEEtran}
\IEEEoverridecommandlockouts
\usepackage{cite}
\usepackage[T1]{fontenc}
\usepackage{graphicx}
\usepackage{amssymb}
\usepackage{amsmath}
\usepackage{amsthm}
\usepackage[caption=false,font=footnotesize]{subfig}
\usepackage{microtype}
\usepackage{balance}
\usepackage{xcolor}
\usepackage{algorithm}
\usepackage{algorithmicx}
\usepackage{algpseudocode}
\usepackage[acronym]{glossaries}
\usepackage{url}
\usepackage{float}
\usepackage{booktabs}
\usepackage{graphicx}
\usepackage{tikz}
\usetikzlibrary{arrows.meta,positioning,shapes.geometric,calc}
\usepackage{makecell}

\algrenewcommand\algorithmicindent{0.7em}%

\newacronym{TW}{TW}{trustworthiness}
\newacronym{CT}{CT}{continuity}
\newacronym{KS}{KS}{Kruskal stress}
\newacronym{MAE}{MAE}{mean absolute error}
\newacronym{CSI}{CSI}{channel-state information}
\newacronym{CC}{CC}{channel charting}
\newacronym{MLP}{MLP}{multilayer perceptron}
\newacronym{ORU}{O-RU}{open RAN radio unit}
\newacronym{COTS}{COTS}{commercial-off-the-shelf}
\newacronym{CAEZ}{CAEZ}{CSI acquisition at ETH Zurich}
\newacronym{WorldVizPPT}{WorldViz PPT}{WorldViz precision position tracking}
\newacronym{AE}{AE}{absolute error}

\newacronym{NPRACH}{NPRACH}{narrowband physical random-access channel}
\newacronym{ToA}{ToA}{time of arrival}
\newacronym{CFO}{CFO}{carrier frequency offset}
\newacronym{NBIoT}{NB-IoT}{narrowband internet of things}
\newacronym{5GNR}{5G NR}{5G New Radio}
\newacronym{3GPP}{3GPP}{3rd Generation Partnership Project}
\newacronym{UMi}{UMi}{urban microcell}
\newacronym{RMSE}{RMSE}{root-mean-square error}
\newacronym{NN}{NN}{neural network}
\newacronym{BS}{BS}{base station}
\newacronym{UE}{UE}{user equipment}
\newacronym{SG}{SG}{symbol group}
\newacronym{CP}{CP}{cyclic prefix}
\newacronym{OFDM}{OFDM}{orthogonal frequency division multiplexing}
\newacronym{FFT}{FFT}{fast Fourier transform}
\newacronym{AWGN}{AWGN}{additive white Gaussian noise}
\newacronym{DFT}{DFT}{discrete Fourier transform}
\newacronym{FNR}{FNR}{false negative rate}
\newacronym{FPR}{FPR}{false positive rate}
\newacronym{RG}{RG}{resource grid}
\newacronym{RE}{RE}{resource element}
\newacronym{SNR}{SNR}{signal-to-noise ratio}
\newacronym{1D}{1D}{one-dimensional}
\newacronym{BCE}{BCE}{binary cross-entropy}
\newacronym{KL}{KL}{Kullback–Leibler}
\newacronym{SGD}{SGD}{stochastic gradient descent}
\newacronym{ppm}{ppm}{parts-per-million}
\newacronym{ICI}{ICI}{inter-carrier interference}
\newacronym{GNN}{GNN}{graph neural network}
\newacronym{BP}{BP}{belief propagation}
\newacronym{FEC}{FEC}{forward error correction}
\newacronym{LDPC}{LDPC}{low-density parity-check}
\newacronym{HDPC}{HDPC}{high-density parity-check}
\newacronym{SCL}{SCL}{successive cancellation list}
\newacronym{SC}{SC}{successive cancellation}
\newacronym{URLLC}{URLLC}{ultra-reliable low-latency communications}
\newacronym{APP}{APP}{a posterior probability}
\newacronym{MIMO}{MIMO}{multiple-input multiple-output}
\newacronym{CNN}{CNN}{convolutional neural network}
\newacronym{BER}{BER}{bit error rate}
\newacronym{BPSK}{BPSK}{binary phase shift keying}
\newacronym{LLR}{LLR}{log-likelihood ratio}
\newacronym{FN}{FN}{factor node}
\newacronym{VN}{VN}{variable node}
\newacronym{CN}{CN}{check node}
\newacronym{MPNN}{MPNN}{message passing neural network}

\newacronym{AI}{AI}{artificial intelligence}
\newacronym{ML}{ML}{machine learning}
\newacronym{SISO}{SISO}{single input single output}
\newacronym{PRB}{PRB}{physical resource block}
\newacronym{PUSCH}{PUSCH}{physical uplink shared channel}
\newacronym{HARQ}{HARQ}{hybrid automatic repeat request}

\newacronym{MUMIMO}{MU-MIMO}{multi-user multiple-input multiple-output}
\newacronym{BICM}{BICM}{bit-interleaved coded modulation}
\newacronym{QAM}{QAM}{quadrature amplitude modulation}
\newacronym{LMMSE}{LMMSE}{linear minimum mean square error}
\newacronym{MMSE}{MMSE}{minimum mean square error}
\newacronym{SIMO}{SIMO}{single-input multiple-output}
\newacronym{CGNN}{CGNN}{convolutional and graph neural network}
\newacronym{BLER}{BLER}{block error rate}
\newacronym{LS}{LS}{least squares}
\newacronym{PE}{PE}{positional encoding}
\newacronym{ReLU}{ReLU}{rectified linear unit}
\newacronym{RB}{RB}{resource block}
\newacronym{CGGNN}{CGGNN}{convolutional graph neural network}
\newacronym{DMRS}{DMRS}{demodulation reference signal}
\newacronym{IoT}{IoT}{internet of things}
\newacronym{ADAM}{ADAM}{adaptive momentum}
\newacronym{TBLER}{TBLER}{transport block error rate}
\newacronym{MCS}{MCS}{modulation and coding scheme}
\newacronym{TDL}{TDL}{tapped delay line}
\newacronym{CDM}{CDM}{code division multiplexing}
\newacronym{FLOP}{FLOP}{floating point operation}
\newacronym{PHY}{PHY}{physical layer}
\newacronym{MAC}{MAC}{media access control}

\newacronym{ULA}{ULA}{uniform linear array}
\newacronym{NRX}{NRX}{neural receiver}
\newacronym{Var-MCS-NRX}{Var-MCS NRX}{variable-MCS NRX}
\newacronym{UL}{UL}{uplink}
\newacronym{DL}{DL}{downlink}
\newacronym{MSE}{MSE}{mean squared error}
\newacronym{CIR}{CIR}{channel impulse response}
\newacronym{iid}{iid}{independent and identically distributed}

\newacronym{RAN}{RAN}{radio access network}
\newacronym{RF}{RF}{radio frequency}
\newacronym{LOS}{LOS}{line-of-sight}
\newacronym{NLOS}{NLOS}{non-line-of-sight}
\newacronym{OTA}{OTA}{over-the-air}
\newacronym{TDD}{TDD}{time-division duplexing}
\newacronym{ARC-OTA}{ARC-OTA}{Aerial RAN CoLab Over-the-Air}
\newacronym{OAI}{OAI}{OpenAirInterface}
\newacronym{RFFI}{RFFI}{radio frequency fingerprint identification}
\newacronym{FH}{FH}{front haul}
\newacronym{GNSS}{GNSS}{global navigation satellite system}
\newacronym{PTP}{PTP}{precision time protocol}
\newacronym{RNTI}{RNTI}{radio network temporary identifier}
\newacronym{ISM}{ISM}{Industrial, Scientific, and Medical}
\newacronym{UAV}{UAV}{unmanned aerial vehicle}
\newacronym{CDF}{CDF}{cumulative distribution function}
\newacronym{CRC}{CRC}{cyclic redundancy check}
\newacronym{SRS}{SRS}{sounding reference signal}
\newacronym{CSI-RS}{CSI-RS}{CSI reference signal}
\usepackage{amssymb}
\usepackage{amsfonts}
\usepackage{mathrsfs}
\usepackage{xspace}
\usepackage{bm}
\usepackage{upgreek}

\newcommand{\safemath}[2]{\newcommand{#1}{\ensuremath{#2}\xspace}}

\safemath{\bma}{\mathbf{a}}
\safemath{\bmb}{\mathbf{b}}
\safemath{\bmc}{\mathbf{c}}
\safemath{\bmd}{\mathbf{d}}
\safemath{\bme}{\mathbf{e}}
\safemath{\bmf}{\mathbf{f}}
\safemath{\bmg}{\mathbf{g}}
\safemath{\bmh}{\mathbf{h}}
\safemath{\bmi}{\mathbf{i}}
\safemath{\bmj}{\mathbf{j}}
\safemath{\bmk}{\mathbf{k}}
\safemath{\bml}{\mathbf{l}}
\safemath{\bmm}{\mathbf{m}}
\safemath{\bmn}{\mathbf{n}}
\safemath{\bmo}{\mathbf{o}}
\safemath{\bmp}{\mathbf{p}}
\safemath{\bmq}{\mathbf{q}}
\safemath{\bmr}{\mathbf{r}}
\safemath{\bms}{\mathbf{s}}
\safemath{\bmt}{\mathbf{t}}
\safemath{\bmu}{\mathbf{u}}
\safemath{\bmv}{\mathbf{v}}
\safemath{\bmw}{\mathbf{w}}
\safemath{\bmx}{\mathbf{x}}
\safemath{\bmy}{\mathbf{y}}
\safemath{\bmz}{\mathbf{z}}
\safemath{\bmzero}{\mathbf{0}}
\safemath{\bmone}{\mathbf{1}}
\safemath{\Bell}{\ensuremath{\boldsymbol\ell}}

\bmdefine{\biad}{a}
\bmdefine{\bibd}{b}
\bmdefine{\bicd}{c}
\bmdefine{\bidd}{d}
\bmdefine{\bied}{e}
\bmdefine{\bifd}{f}
\bmdefine{\bigd}{g}
\bmdefine{\bihd}{h}
\bmdefine{\biid}{i}
\bmdefine{\bijd}{j}
\bmdefine{\bikd}{k}
\bmdefine{\bild}{l}
\bmdefine{\bimd}{m}
\bmdefine{\bind}{n}
\bmdefine{\biod}{o}
\bmdefine{\bipd}{p}
\bmdefine{\biqd}{q}
\bmdefine{\bird}{r}
\bmdefine{\bisd}{s}
\bmdefine{\bitd}{t}
\bmdefine{\biud}{u}
\bmdefine{\bivd}{v}
\bmdefine{\biwd}{w}
\bmdefine{\bixd}{x}
\bmdefine{\biyd}{y}
\bmdefine{\bizd}{z}

\bmdefine{\bixid}{\xi}
\bmdefine{\bilambdad}{\lambda}
\bmdefine{\bimud}{\mu}
\bmdefine{\bithetad}{\theta}
\bmdefine{\biphid}{\phi}
\bmdefine{\bideltad}{\delta}

\safemath{\bmia}{\biad}
\safemath{\bmib}{\bibd}
\safemath{\bmic}{\bicd}
\safemath{\bmid}{\bidd}
\safemath{\bmie}{\bied}
\safemath{\bmif}{\bifd}
\safemath{\bmig}{\bigd}
\safemath{\bmih}{\bihd}
\safemath{\bmii}{\biid}
\safemath{\bmij}{\bijd}
\safemath{\bmik}{\bikd}
\safemath{\bmil}{\bild}
\safemath{\bmim}{\bimd}
\safemath{\bmin}{\bind}
\safemath{\bmio}{\biod}
\safemath{\bmip}{\bipd}
\safemath{\bmiq}{\biqd}
\safemath{\bmir}{\bird}
\safemath{\bmis}{\bisd}
\safemath{\bmit}{\bitd}
\safemath{\bmiu}{\biud}
\safemath{\bmiv}{\bivd}
\safemath{\bmiw}{\biwd}
\safemath{\bmix}{\bixd}
\safemath{\bmiy}{\biyd}
\safemath{\bmiz}{\bizd}

\safemath{\bmxi}{\bixid}
\safemath{\bmlambda}{\bilambdad}
\safemath{\bmmu}{\bimud}
\safemath{\bmtheta}{\bithetad}
\safemath{\bmphi}{\biphid}
\safemath{\bmdelta}{\bideltad}

\safemath{\bA}{\mathbf{A}}
\safemath{\bB}{\mathbf{B}}
\safemath{\bC}{\mathbf{C}}
\safemath{\bD}{\mathbf{D}}
\safemath{\bE}{\mathbf{E}}
\safemath{\bF}{\mathbf{F}}
\safemath{\bG}{\mathbf{G}}
\safemath{\bH}{\mathbf{H}}
\safemath{\bI}{\mathbf{I}}
\safemath{\bJ}{\mathbf{J}}
\safemath{\bK}{\mathbf{K}}
\safemath{\bL}{\mathbf{L}}
\safemath{\bM}{\mathbf{M}}
\safemath{\bN}{\mathbf{N}}
\safemath{\bO}{\mathbf{O}}
\safemath{\bP}{\mathbf{P}}
\safemath{\bQ}{\mathbf{Q}}
\safemath{\bR}{\mathbf{R}}
\safemath{\bS}{\mathbf{S}}
\safemath{\bT}{\mathbf{T}}
\safemath{\bU}{\mathbf{U}}
\safemath{\bV}{\mathbf{V}}
\safemath{\bW}{\mathbf{W}}
\safemath{\bX}{\mathbf{X}}
\safemath{\bY}{\mathbf{Y}}
\safemath{\bZ}{\mathbf{Z}}

\safemath{\bZero}{\mathbf{0}}
\safemath{\bOne}{\mathbf{1}}
\safemath{\bDelta}{\mathbf{\Delta}}
\safemath{\bLambda}{\mathbf{\UpLambda}}
\safemath{\bPhi}{\mathbf{\Upphi}}
\safemath{\bSigma}{\mathbf{\Upsigma}}
\safemath{\bOmega}{\mathbf{\Upomega}}
\safemath{\bTheta}{\mathbf{\Uptheta}}

\bmdefine{\biAd}{A}
\bmdefine{\biBd}{B}
\bmdefine{\biCd}{C}
\bmdefine{\biDd}{D}
\bmdefine{\biEd}{E}
\bmdefine{\biFd}{F}
\bmdefine{\biGd}{G}
\bmdefine{\biHd}{H}
\bmdefine{\biId}{I}
\bmdefine{\biJd}{J}
\bmdefine{\biKd}{K}
\bmdefine{\biLd}{L}
\bmdefine{\biMd}{M}
\bmdefine{\biOd}{N}
\bmdefine{\biPd}{O}
\bmdefine{\biQd}{P}
\bmdefine{\biRd}{R}
\bmdefine{\biSd}{S}
\bmdefine{\biTd}{T}
\bmdefine{\biUd}{U}
\bmdefine{\biVd}{V}
\bmdefine{\biWd}{W}
\bmdefine{\biXd}{X}
\bmdefine{\biYd}{Y}
\bmdefine{\biZd}{Z}

\bmdefine{\biDelta}{\Delta}
\bmdefine{\biLambda}{\Lambda}
\bmdefine{\biPhi}{\Phi}
\bmdefine{\biSigma}{\Sigma}
\bmdefine{\biOmega}{\Omega}
\bmdefine{\biTheta}{\Theta}

\safemath{\bimA}{\biAd}
\safemath{\bimB}{\biBd}
\safemath{\bimC}{\biCd}
\safemath{\bimD}{\biDd}
\safemath{\bimE}{\biEd}
\safemath{\bimF}{\biFd}
\safemath{\bimG}{\biGd}
\safemath{\bimH}{\biHd}
\safemath{\bimI}{\biId}
\safemath{\bimJ}{\biJd}
\safemath{\bimK}{\biKd}
\safemath{\bimL}{\biLd}
\safemath{\bimM}{\biMd}
\safemath{\bimN}{\biNd}
\safemath{\bimO}{\biOd}
\safemath{\bimP}{\biPd}
\safemath{\bimQ}{\biQd}
\safemath{\bimR}{\biRd}
\safemath{\bimS}{\biSd}
\safemath{\bimT}{\biTd}
\safemath{\bimU}{\biUd}
\safemath{\bimV}{\biVd}
\safemath{\bimW}{\biWd}
\safemath{\bimX}{\biXd}
\safemath{\bimY}{\biYd}
\safemath{\bimZ}{\biZd}

\safemath{\bimDelta}{\biDelta}
\safemath{\bimLambda}{\biLambda}
\safemath{\bimPhi}{\biPhi}
\safemath{\bimSigma}{\biSigma}
\safemath{\bimOmega}{\biOmega}
\safemath{\bimTheta}{\biTheta}

\safemath{\setA}{\mathcal{A}}
\safemath{\setB}{\mathcal{B}}
\safemath{\setC}{\mathcal{C}}
\safemath{\setD}{\mathcal{D}}
\safemath{\setE}{\mathcal{E}}
\safemath{\setF}{\mathcal{F}}
\safemath{\setG}{\mathcal{G}}
\safemath{\setH}{\mathcal{H}}
\safemath{\setI}{\mathcal{I}}
\safemath{\setJ}{\mathcal{J}}
\safemath{\setK}{\mathcal{K}}
\safemath{\setL}{\mathcal{L}}
\safemath{\setM}{\mathcal{M}}
\safemath{\setN}{\mathcal{N}}
\safemath{\setO}{\mathcal{O}}
\safemath{\setP}{\mathcal{P}}
\safemath{\setQ}{\mathcal{Q}}
\safemath{\setR}{\mathcal{R}}
\safemath{\setS}{\mathcal{S}}
\safemath{\setT}{\mathcal{T}}
\safemath{\setU}{\mathcal{U}}
\safemath{\setV}{\mathcal{V}}
\safemath{\setW}{\mathcal{W}}
\safemath{\setX}{\mathcal{X}}
\safemath{\setY}{\mathcal{Y}}
\safemath{\setZ}{\mathcal{Z}}
\safemath{\emptySet}{\varnothing}

\safemath{\colA}{\mathscr{A}}
\safemath{\colB}{\mathscr{B}}
\safemath{\colC}{\mathscr{C}}
\safemath{\colD}{\mathscr{D}}
\safemath{\colE}{\mathscr{E}}
\safemath{\colF}{\mathscr{F}}
\safemath{\colG}{\mathscr{G}}
\safemath{\colH}{\mathscr{H}}
\safemath{\colI}{\mathscr{I}}
\safemath{\colJ}{\mathscr{J}}
\safemath{\colK}{\mathscr{K}}
\safemath{\colL}{\mathscr{L}}
\safemath{\colM}{\mathscr{M}}
\safemath{\colN}{\mathscr{N}}
\safemath{\colO}{\mathscr{O}}
\safemath{\colP}{\mathscr{P}}
\safemath{\colQ}{\mathscr{Q}}
\safemath{\colR}{\mathscr{R}}
\safemath{\colS}{\mathscr{S}}
\safemath{\colT}{\mathscr{T}}
\safemath{\colU}{\mathscr{U}}
\safemath{\colV}{\mathscr{V}}
\safemath{\colW}{\mathscr{W}}
\safemath{\colX}{\mathscr{X}}
\safemath{\colY}{\mathscr{Y}}
\safemath{\colZ}{\mathscr{Z}}

\safemath{\opA}{\mathbb{A}}
\safemath{\opB}{\mathbb{B}}
\safemath{\opC}{\mathbb{C}}
\safemath{\opD}{\mathbb{D}}
\safemath{\opE}{\mathbb{E}}
\safemath{\opF}{\mathbb{F}}
\safemath{\opG}{\mathbb{G}}
\safemath{\opH}{\mathbb{H}}
\safemath{\opI}{\mathbb{I}}
\safemath{\opJ}{\mathbb{J}}
\safemath{\opK}{\mathbb{K}}
\safemath{\opL}{\mathbb{L}}
\safemath{\opM}{\mathbb{M}}
\safemath{\opN}{\mathbb{N}}
\safemath{\opO}{\mathbb{O}}
\safemath{\opP}{\mathbb{P}}
\safemath{\opQ}{\mathbb{Q}}
\safemath{\opR}{\mathbb{R}}
\safemath{\opS}{\mathbb{S}}
\safemath{\opT}{\mathbb{T}}
\safemath{\opU}{\mathbb{U}}
\safemath{\opV}{\mathbb{V}}
\safemath{\opW}{\mathbb{W}}
\safemath{\opX}{\mathbb{X}}
\safemath{\opY}{\mathbb{Y}}
\safemath{\opZ}{\mathbb{Z}}
\safemath{\opZero}{\mathbb{O}}
\safemath{\identityop}{\opI}

\safemath{\veca}{\bma}
\safemath{\vecb}{\bmb}
\safemath{\vecc}{\bmc}
\safemath{\vecd}{\bmd}
\safemath{\vece}{\bme}
\safemath{\vecf}{\bmf}
\safemath{\vecg}{\bmg}
\safemath{\vech}{\bmh}
\safemath{\veci}{\bmi}
\safemath{\vecj}{\bmj}
\safemath{\veck}{\bmk}
\safemath{\vecl}{\bml}
\safemath{\vecm}{\bmm}
\safemath{\vecn}{\bmn}
\safemath{\veco}{\bmo}
\safemath{\vecp}{\bmp}
\safemath{\vecq}{\bmq}
\safemath{\vecr}{\bmr}
\safemath{\vecs}{\bms}
\safemath{\vect}{\bmt}
\safemath{\vecu}{\bmu}
\safemath{\vecv}{\bmv}
\safemath{\vecw}{\bmw}
\safemath{\vecx}{\bmx}
\safemath{\vecy}{\bmy}
\safemath{\vecz}{\bmz}

\safemath{\veczero}{\bmzero}
\safemath{\vecone}{\bmone}
\safemath{\vecxi}{\bmxi}
\safemath{\veclambda}{\bmlambda}
\safemath{\vecmu}{\bmmu}
\safemath{\vectheta}{\bmtheta}
\safemath{\vecphi}{\bmphi}
\safemath{\vecdelta}{\bmdelta}

\safemath{\matA}{\bA}
\safemath{\matB}{\bB}
\safemath{\matC}{\bC}
\safemath{\matD}{\bD}
\safemath{\matE}{\bE}
\safemath{\matF}{\bF}
\safemath{\matG}{\bG}
\safemath{\matH}{\bH}
\safemath{\matI}{\bI}
\safemath{\matJ}{\bJ}
\safemath{\matK}{\bK}
\safemath{\matL}{\bL}
\safemath{\matM}{\bM}
\safemath{\matN}{\bN}
\safemath{\matO}{\bO}
\safemath{\matP}{\bP}
\safemath{\matQ}{\bQ}
\safemath{\matR}{\bR}
\safemath{\matS}{\bS}
\safemath{\matT}{\bT}
\safemath{\matU}{\bU}
\safemath{\matV}{\bV}
\safemath{\matW}{\bW}
\safemath{\matX}{\bX}
\safemath{\matY}{\bY}
\safemath{\matZ}{\bZ}
\safemath{\matzero}{\bmzero}

\safemath{\matDelta}{\bDelta}
\safemath{\matLambda}{\bLambda}
\safemath{\matPhi}{\bPhi}
\safemath{\matSigma}{\bSigma}
\safemath{\matOmega}{\bOmega}
\safemath{\matTheta}{\bTheta}

\safemath{\matidentity}{\matI}
\safemath{\matone}{\matO}

\safemath{\rnda}{A}
\safemath{\rndb}{B}
\safemath{\rndc}{C}
\safemath{\rndd}{D}
\safemath{\rnde}{E}
\safemath{\rndf}{F}
\safemath{\rndg}{G}
\safemath{\rndh}{H}
\safemath{\rndi}{I}
\safemath{\rndj}{J}
\safemath{\rndk}{K}
\safemath{\rndl}{L}
\safemath{\rndm}{M}
\safemath{\rndn}{N}
\safemath{\rndo}{O}
\safemath{\rndp}{P}
\safemath{\rndq}{Q}
\safemath{\rndr}{R}
\safemath{\rnds}{S}
\safemath{\rndt}{T}
\safemath{\rndu}{U}
\safemath{\rndv}{V}
\safemath{\rndw}{W}
\safemath{\rndx}{X}
\safemath{\rndy}{Y}
\safemath{\rndz}{Z}

\safemath{\rveca}{\bimA}
\safemath{\rvecb}{\bimB}
\safemath{\rvecc}{\bimC}
\safemath{\rvecd}{\bimD}
\safemath{\rvece}{\bimE}
\safemath{\rvecf}{\bimF}
\safemath{\rvecg}{\bimG}
\safemath{\rvech}{\bimH}
\safemath{\rveci}{\bimI}
\safemath{\rvecj}{\bimJ}
\safemath{\rveck}{\bimK}
\safemath{\rvecl}{\bimL}
\safemath{\rvecm}{\bimM}
\safemath{\rvecn}{\bimN}
\safemath{\rveco}{\bomO}
\safemath{\rvecp}{\bimP}
\safemath{\rvecq}{\bimQ}
\safemath{\rvecr}{\bimR}
\safemath{\rvecs}{\bimS}
\safemath{\rvect}{\bimT}
\safemath{\rvecu}{\bimU}
\safemath{\rvecv}{\bimV}
\safemath{\rvecw}{\bimW}
\safemath{\rvecx}{\bimX}
\safemath{\rvecy}{\bimY}
\safemath{\rvecz}{\bimZ}

\safemath{\rvecxi}{\bmxi}
\safemath{\rveclambda}{\bmlambda}
\safemath{\rvecmu}{\bmmu}
\safemath{\rvectheta}{\bmtheta}
\safemath{\rvecphi}{\bmphi}

\safemath{\rmatA}{\bimA}
\safemath{\rmatB}{\bimB}
\safemath{\rmatC}{\bimC}
\safemath{\rmatD}{\bimD}
\safemath{\rmatE}{\bimE}
\safemath{\rmatF}{\bimF}
\safemath{\rmatG}{\bimG}
\safemath{\rmatH}{\bimH}
\safemath{\rmatI}{\bimI}
\safemath{\rmatJ}{\bimJ}
\safemath{\rmatK}{\bimK}
\safemath{\rmatL}{\bimL}
\safemath{\rmatM}{\bimM}
\safemath{\rmatN}{\bimN}
\safemath{\rmatO}{\bimO}
\safemath{\rmatP}{\bimP}
\safemath{\rmatQ}{\bimQ}
\safemath{\rmatR}{\bimR}
\safemath{\rmatS}{\bimS}
\safemath{\rmatT}{\bimT}
\safemath{\rmatU}{\bimU}
\safemath{\rmatV}{\bimV}
\safemath{\rmatW}{\bimW}
\safemath{\rmatX}{\bimX}
\safemath{\rmatY}{\bimY}
\safemath{\rmatZ}{\bimZ}

\safemath{\rmatDelta}{\bimDelta}
\safemath{\rmatLambda}{\bimLambda}
\safemath{\rmatPhi}{\bimPhi}
\safemath{\rmatSigma}{\bimSigma}
\safemath{\rmatOmega}{\bimOmega}
\safemath{\rmatTheta}{\bimTheta}

\usepackage{amssymb}
\usepackage{amsfonts}
\usepackage{mathrsfs}
\usepackage{xspace}
\usepackage{bm}
\usepackage{fancyref}
\usepackage{textcomp}

\usepackage{multirow}
\usepackage{stmaryrd}

\usepackage{booktabs}

\newenvironment{textbmatrix}{	\setlength{\arraycolsep}{2.5pt}%
								\left[\begin{matrix}}{\end{matrix}\right]%
								\raisebox{0.08ex}{\vphantom{M}}}

\def\be{\begin{equation}}
\def\ee{\end{equation}}
\def\een{\nonumber \end{equation}}
\def\mat{\begin{bmatrix}}
\def\emat{\end{bmatrix}}
\def\btm{\begin{textbmatrix}}
\def\etm{\end{textbmatrix}}

\def\ba#1\ea{\begin{align}#1\end{align}}
\def\bas#1\eas{\begin{align*}#1\end{align*}}
\def\bs#1\es{\begin{split}#1\end{split}}
\def\bg#1\eg{\begin{gather}#1\end{gather}}
\def\bml#1\eml{\begin{multline}#1\end{multline}}
\def\bi#1\ei{\begin{itemize}#1\end{itemize}}

\newcommand{\lefto}{\mathopen{}\left}

\newcommand{\vecnorm}[1]{\lefto\lVert#1\right\rVert}		

\safemath{\dirac}{\delta}					
\safemath{\krond}{\dirac}					

\safemath{\upto}{\uparrow}
\safemath{\downto}{\downarrow}
\safemath{\iu}{j}							
\safemath{\ev}{\lambda}						
\safemath{\hilseqspace}{l^{2}}				
\newcommand{\banachfunspace}[1]{\setL^{#1}}	
\safemath{\hilfunspace}{\banachfunspace{2}}	

\safemath{\SNR}{\textit{SNR}} 				
\safemath{\PAR}{\textit{PAR}} 				
\safemath{\No}{N_0}							
\safemath{\Es}{E_s}							
\safemath{\Eb}{E_b}							
\safemath{\EbNo}{\frac{\Eb}{\No}}
\safemath{\EsNo}{\frac{\Es}{\No}}

\DeclareMathOperator{\CHop}{\ensuremath{\opH}} 
\safemath{\tvir}{\rndh_{\CHop}}				
\safemath{\tvtf}{\rndl_{\CHop}}				
\safemath{\spf}{\rnds_{\CHop}}				
\safemath{\bff}{H_{\CHop}}					

\safemath{\ircf}{r_{h}}						
\safemath{\tftvcf}{r_{s}}					
\safemath{\tfcf}{r_{l}}						
\safemath{\bfcf}{r_{H}}						

\safemath{\tcorr}{c_h}						
\safemath{\scf}{c_{s}}						
\safemath{\tfcorr}{c_{l}}					
\safemath{\fcorr}{c_{H}}						

\safemath{\mi}{I}							
\safemath{\capacity}{C}						

\safemath{\normal}{\mathcal{N}}			
\safemath{\jpg}{\mathcal{CN}}			
\safemath{\mchain}{\leftrightarrow}		

\safemath{\dB}{\,\mathrm{dB}}
\safemath{\dBm}{\,\mathrm{dBm}}
\safemath{\Hz}{\,\mathrm{Hz}}
\safemath{\kHz}{\,\mathrm{kHz}}
\safemath{\MHz}{\,\mathrm{MHz}}
\safemath{\GHz}{\,\mathrm{GHz}}
\safemath{\s}{\,\mathrm{s}}
\safemath{\ms}{\,\mathrm{ms}}
\safemath{\mus}{\,\mathrm{\text{\textmu}s}}
\safemath{\ns}{\,\mathrm{ns}}
\safemath{\ps}{\,\mathrm{ps}}
\safemath{\meter}{\,\mathrm{m}}
\safemath{\mm}{\,\mathrm{mm}}
\safemath{\cm}{\,\mathrm{cm}}
\safemath{\m}{\,\mathrm{m}}
\safemath{\W}{\,\mathrm{W}}
\safemath{\mW}{\, \mathrm{mW}}
\safemath{\J}{\,\mathrm{J}}
\safemath{\K}{\,\mathrm{K}}
\safemath{\bit}{\,\mathrm{bit}}
\safemath{\nat}{\,\mathrm{nat}}

\safemath{\define}{\triangleq}			

\safemath{\equivalent}{\sim}
\safemath{\distas}{\sim}					
\safemath{\sdiff}{\Delta}				

\safemath{\reals}{\mathbb{R}}
\safemath{\positivereals}{\reals_{+}}
\safemath{\integers}{\mathbb{Z}}
\safemath{\posint}{\integers_{+}}
\safemath{\naturals}{\mathbb{N}}
\safemath{\posnaturals}{\naturals_{+}}
\safemath{\complexset}{\mathbb{C}}
\safemath{\rationals}{\mathbb{Q}}

\newcommand*{\fancyrefapplabelprefix}{app}		
\newcommand*{\fancyrefthmlabelprefix}{thm}		
\newcommand*{\fancyreflemlabelprefix}{lem}		
\newcommand*{\fancyrefcorlabelprefix}{cor}		
\newcommand*{\fancyrefdeflabelprefix}{def}		
\newcommand*{\fancyrefproplabelprefix}{prop}		
\newcommand*{\fancyrefexmpllabelprefix}{exmpl}
\newcommand*{\fancyrefalglabelprefix}{alg}		
\newcommand*{\fancyreftbllabelprefix}{tbl}		

\frefformat{vario}{\fancyrefseclabelprefix}{Sec.~#1}
\frefformat{vario}{\fancyrefthmlabelprefix}{Thm.~#1}
\frefformat{vario}{\fancyreftbllabelprefix}{Tbl.~#1}
\frefformat{vario}{\fancyreflemlabelprefix}{Lem.~#1}
\frefformat{vario}{\fancyrefcorlabelprefix}{Corr.~#1}
\frefformat{vario}{\fancyrefdeflabelprefix}{Def.~#1}
\frefformat{vario}{\fancyreffiglabelprefix}{Fig.~#1}
\frefformat{vario}{\fancyrefapplabelprefix}{App.~#1}
\frefformat{vario}{\fancyrefeqlabelprefix}{(#1)}
\frefformat{vario}{\fancyrefproplabelprefix}{Prop.~#1}
\frefformat{vario}{\fancyrefexmpllabelprefix}{Ex.~#1}
\frefformat{vario}{\fancyrefalglabelprefix}{Alg.~#1}

\DeclareMathOperator{\atantwo}{atan2}

\safemath{\dictab}{[\,\dicta\,\,\dictb\,]}

\safemath{\ysig}{\bmy}
\safemath{\ysighat}{\hat{\ysig}}
\safemath{\ysigdim}{M}
\safemath{\xsig}{\bmx}
\safemath{\xsigdim}{N}
\safemath{\nx}{n_x}
\safemath{\zsig}{\bmz}
\safemath{\zsigdim}{\ysigdim}
\safemath{\rsig}{\bmr}
\safemath{\Adict}{\bA}
\safemath{\Adicttilde}{\widetilde{\Adict}}
\safemath{\Adictdim}{\outputdim\times\xsigdim}
\safemath{\avec}{\bma}
\safemath{\avectilde}{\tilde{\avec}}
\safemath{\Bdict}{\bB}
\safemath{\Bdicttilde}{\widetilde{\Bdict}}
\safemath{\Cdict}{\bC}
\safemath{\cvec}{\bmc}
\safemath{\Ddict}{\bD}
\safemath{\Ddictdim}{\ysigdim\times\xsigdim}
\safemath{\dvec}{\bmd}
\safemath{\Ddicttilde}{\widetilde{\bD}}
\safemath{\Bonb}{\bB}
\safemath{\bvec}{\bmb}
\safemath{\Bonbdim}{\ysigdim\times\ysigdim}
\safemath{\noise}{\bmn}
\safemath{\noisedim}{\ysigim}
\safemath{\err}{\bme}
\safemath{\errdim}{\ysigdim}
\safemath{\errset}{\setE}
\safemath{\nerr}{n_e}
\safemath{\delop}{\bP_\errset}
\safemath{\delopc}{\bP_{{\errset}^c}}

\safemath{\cplxi}{\imath}
\safemath{\cplxj}{\jmath}

\safemath{\dict}{\matD}
\safemath{\inputdim}{N}		
\safemath{\outputdim}{M}		
\safemath{\sparsity}{S}	
\safemath{\inputdimA}{{N_a}}	
\safemath{\inputdimB}{{N_b}}	
\safemath{\elemA}{{n_a}}	
\safemath{\elemB}{{n_b}}	
\safemath{\resA}{\matR_a}	
\safemath{\resB}{\matR_b}	
\safemath{\subD}{\matS} 
\safemath{\subA}{\matS_a} 
\safemath{\subB}{\matS_b} 
\safemath{\dicta}{\matA} 	
\safemath{\dictb}{\matB} 	
\safemath{\hollowS}{H}
\safemath{\hollowA}{H_a}
\safemath{\hollowB}{H_b}
\safemath{\cross}{Z}
\safemath{\coh}{\mu_d}			
\safemath{\coha}{\mu_a}			
\safemath{\cohb}{\mu_b}			
\safemath{\mubs}{\nu}	
\safemath{\cohm}{\mu_m} 
\safemath{\dictset}{\setD}	
\safemath{\dictsetp}{\dictset(\coh,\coha,\cohb)}	
\safemath{\dictsetgen}{\dictset_\text{gen}}
\safemath{\dictsetgenp}{\dictsetgen(\coh)}
\safemath{\dictsetonb}{\dictset_\text{onb}}
\safemath{\dictsetonbp}{\dictsetonb(\coh)}

\safemath{\leftside}{U}
\safemath{\rightsideA}{R_a}
\safemath{\rightsideB}{R_b}

\safemath{\indexS}{\setI_S} 

\safemath{\na}{n_a}			
\safemath{\nb}{n_b}			
\safemath{\coeffa}{p_i}	
\safemath{\coeffb}{q_j}	
\safemath{\seta}{\setP}		
\safemath{\setb}{\setQ}     
\safemath{\setw}{\setW}	
\safemath{\setz}{\setZ}	
\safemath{\cola}{\veca}		
\safemath{\colb}{\vecb}		
\safemath{\cold}{\vecd}		
\safemath{\inputvec}{\vecx} 	
\safemath{\error}{\vece}	
\safemath{\noiseout}{\vecz} 	
\safemath{\inputvecel}{x}
\safemath{\inputveca}{\vecx_a}
\safemath{\inputvecb}{\vecx_b}
\safemath{\outputvec}{\vecy}	
\safemath{\lambdamin}{\lambda_{\mathrm{min}}}

\safemath{\elltwo}{\ell_2}
\safemath{\ellone}{\ell_1}
\safemath{\ellzero}{\ell_0}
\safemath{\ellinf}{\ell_\infty}
\safemath{\ellinftilde}{\ell_{\widetilde\infty}}
\safemath{\licard}{Z(\coh,\coha,\cohb)}
\safemath{\xsol}{\hat{x}}
\safemath{\xbord}{x_b}		
\safemath{\xstat}{x_s}		
\safemath{\xstatLone}{\tilde{x}_s}
\safemath{\order}{\mathcal{O}} 
\safemath{\scales}{\Theta} 
\safemath{\ones}{\mathbf{1}} 
\safemath{\zeroes}{\mathbf{0}} 
\safemath{\thlone}{\kappa(\coh,\cohb)} 
\safemath{\constoneA}{\delta} 
\safemath{\constoneB}{\epsilon} 
\safemath{\nlarge}{L}				   
\safemath{\sumlarge}{S_\nlarge}
\safemath{\maxlarger}{P_\nlarge}	   
\safemath{\Pzero}{\textrm{P0}}	
\safemath{\Pone}{\textrm{P1}}
\safemath{\vecfir}{\vecw}			 
\safemath{\vecsec}{\vecz}
\safemath{\elvecfir}{w}              
\safemath{\elvecsec}{z}				 
\safemath{\nlargefir}{n}
\safemath{\normout}{\gamma}
\safemath{\auxfun}{h}
\safemath{\supp}{\textrm{supp}}

\safemath{\indexa}{\ell}
\safemath{\indexb}{r}
\safemath{\indexc}{i}
\safemath{\indexd}{j}

\safemath{\project}{P}

\safemath{\firstslotset}{\setU_1}  
\safemath{\secondslotset}{\setU_2} 
\safemath{\randomset}{\setS} 

\safemath{\Tran}{\textnormal{T}}
\safemath{\Herm}{\textnormal{H}} 

\definecolor{myred}{HTML}{9E292B}
\definecolor{myblue}{HTML}{235787}

\newcommand*{\fancyreflstlabelprefix}{lst}
\fancyrefaddcaptions{english}{%
  \providecommand*{\freflstname}{Listing}%
}
\frefformat{vario}{\fancyreflstlabelprefix}{%
  \freflstname\fancyrefdefaultspacing#1#3%
}

\begin{document}
\title{Channel Charting for Position and Orientation}
\author{Daniel Richner, Reinhard Wiesmayr, Frederik Zumegen, and Christoph Studer\\[0.25cm]
\em ETH Z\"urich, Switzerland; e-mail: wiesmayr@iis.ee.ethz.ch\\[-.5cm]
\thanks{This work was supported in part by the Swiss National Science Foundation (SNSF) grant 200021\_207314 and by CHIST-ERA grant for the project CHASER (CHIST-ERA-22-WAI-01) through the SNSF grant 20CH21\_218704. We acknowledge NVIDIA for its sponsorship of this research. We also acknowledge Ant\'onio Maia Barros for his support in our experimental analysis on uplink transmit precoding configurations.}
}

\maketitle

\bstctlcite{BSTcontrol}

\begin{abstract}
Channel charting (CC) in real-world coordinates is a recently proposed self-supervised machine learning method that maps high-dimensional channel state information (CSI) to user equipment (UE) position. In this paper, we extend CC to also estimate UE orientation, which can further assist tasks such as beamfinding, precoding, and beam- and cell-assignment. To this end, we propose a novel orientation triplet loss that accounts for angle periodicity and an alignment loss that embeds estimated orientations in real-world coordinates in a self-supervised fashion. Using real-world CSI measurements from a standard-compliant 5G NR system, we demonstrate that the proposed method achieves position and orientation estimation accuracy close to that of supervised approaches trained with ground-truth labels.
\end{abstract}

\section{Introduction}

Future wireless communication systems are expected to incorporate \gls{UE} localization capabilities \cite{6g_localization_sensing}. \Gls{CC} is a self-supervised pseudo-positioning method that maps high-dimensional \gls{CSI} to a low-dimensional representation associated with \gls{UE} position without requiring ground-truth labels during training~\cite{studer2018channelcharting}. The recent approach in~\cite{taner2025chartingrealworld} further enables positioning in real-world coordinates. Existing \gls{CC} methods, however, focus solely on position estimation and ignore inherent \gls{UE} orientation dependence of \gls{CSI}~\cite{wiesmayr2025csi,stephan2024angledelay,stephan2025threedim}.

Information on \gls{UE} orientation promises significant benefits for wireless networks. In particular, orientation information can assist beamfinding, precoding, beam- and cell-assignment, and \gls{UE} trajectory prediction. Furthermore, when combined with information on the \gls{UE} antenna placement, orientation estimates can indicate the heading (i.e., \gls{UE} movement direction) of  vehicles on the ground or in the air that carry the \gls{UE}.

\subsection{Contributions}

\begin{figure}[t]
    \centering
    \includegraphics[width=0.99\linewidth]{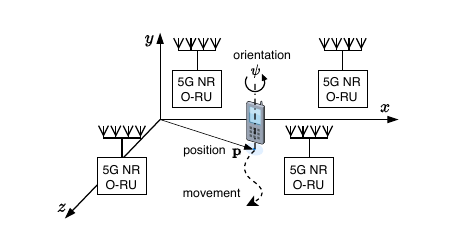}
    \vspace{-0.8cm}
    \caption{Moving \gls{UE} in multi-cell 5G NR system used for position and orientation estimation. We propose a neural-network-based single-shot method that estimates \gls{UE} position $\bmp$ \emph{and} yaw angle $\psi$ with self-supervised training.}
    \label{fig:problem_visualization}
\end{figure}

We propose a new \gls{CC} pipeline that maps a single \gls{CSI} feature not only to a \gls{UE} position estimate but also to an orientation estimate, as illustrated in \fref{fig:problem_visualization}. Our training pipeline builds on triplet-based \gls{CC} \cite{ferrand2022tripletbased} in real-world coordinates \cite{taner2025chartingrealworld} and extends it with (i) a novel orientation triplet loss and (ii) a novel alignment loss. Using a new 5G NR measurement dataset, we first demonstrate that \gls{CSI} depends strongly on the \gls{UE} orientation. We then evaluate the proposed method on measured \gls{CSI} from a \gls{COTS} \gls{UE} using the CAEZ-5G-OUTDOOR dataset presented in~\cite{wiesmayr2025csi}.

\subsection{Relevant Prior Art}

The existing literature largely reports \gls{CC} only for positioning, where most work focuses on two dimensions (2D) \cite{wiesmayr2025csi, taner2025chartingrealworld}; the recent work in~\cite{stephan2025threedim} also studies \gls{CC} in three dimensions (3D). In contrast, we propose a novel \gls{CC}-based positioning pipeline that also provides orientation information from single-shot \gls{CSI} measurements (i.e., by mapping one \gls{CSI} feature to position and orientation estimates). While one could extract the heading of a moving \gls{UE} from its motion trajectory using multi-shot \gls{CC} (i.e., by mapping a time-series of \gls{CSI} features to a trajectory), we extract orientation information from a single \gls{CSI} feature. 

Recent work in~\cite{zhao2026jointlocalizationorientationtriplebeam} proposes supervised position and heading estimation from delay-Doppler features of synthetic \gls{CSI} data, so that training requires ground-truth labels. In contrast, our method relies on self-supervised training. Furthermore, we evaluate our method using real-world \gls{CSI} measurements in a standard-compliant \gls{5GNR} system.

\section{CSI Depends on Orientation}
\label{sec:csi_orientation}

As already mentioned in \cite[Sec. II]{stephan2024angledelay}, \gls{CSI} depends on \gls{UE} orientation. We now experimentally confirm this statement using measured \gls{CSI} from the 5G NR \gls{PUSCH} that leverages \glspl{DMRS} transmitted by \gls{COTS} \glspl{UE}. In general, \glspl{UE} do not transmit \gls{RF} signals in an ideal omnidirectional pattern, but instead transmit with an orientation-dependent antenna radiation pattern. To characterize this angular dependency, we utilize a \gls{COTS} \gls{UE} that transmits \gls{PUSCH} with one spatial stream (layer), and we measure wideband \gls{CSI} with four distributed multi-antenna \gls{COTS} \glspl{ORU}. We observe that the measured \gls{CSI} varies with \gls{UE} orientation, and that this angular dependency is reproducible for repeated measurements at the same location. 

\subsection{5G NR Testbed at ETH Zurich}

We use the 5G NR testbed at ETH Zurich \cite{wiesmayr2025csi}, which is based on the NVIDIA Aerial Testbed (ATB)~\cite{nvidia_ATB_2026}. This testbed is a full-stack software-defined 5G NR system with \gls{COTS} \glspl{UE} and four \gls{COTS} \glspl{ORU}, in which one \gls{ORU} is used for communication and the others are passive listeners. We use the same setup, configuration, and acquisition pipeline as detailed in~\cite{wiesmayr2025csi}. The \gls{CSI} measurements are acquired in the Swiss private 5G band with a bandwidth of $100$\,MHz centered at $3.45$\,GHz and $30$\,kHz subcarrier spacing. Our testbed configuration with enabled \gls{CSI-RS} and disabled \gls{SRS} leads to static uplink precoding with only the \gls{UE}'s primary transmit antenna active. This configuration ensures that, for a fixed \gls{UE} position, any change in \gls{CSI} during \gls{UE} rotation is caused by the \gls{UE}'s orientation.

\subsection{CAEZ-5G-ORIENTATION Dataset}

We measured a new 5G NR \gls{CSI} dataset called CAEZ-5G-ORIENTATION, in which each \gls{CSI} sample is labeled with the ground-truth \gls{UE} position and orientation, analogous to the existing \gls{CAEZ} datasets~\cite{wiesmayr2025csi}.
The key dataset properties are listed in \fref{tbl:datasets}. We use a measurement environment similar to that of the CAEZ-5G-INDOOR dataset in \cite[Fig.~3c]{wiesmayr2025csi}, where the \glspl{ORU} were placed at the corners of a small laboratory room.
We obtain ground-truth position and orientation labels using the same \gls{WorldVizPPT} system as in \cite{wiesmayr2025csi}.

\begin{figure}[tb]
    \centering
    \includegraphics[width=0.5\linewidth]{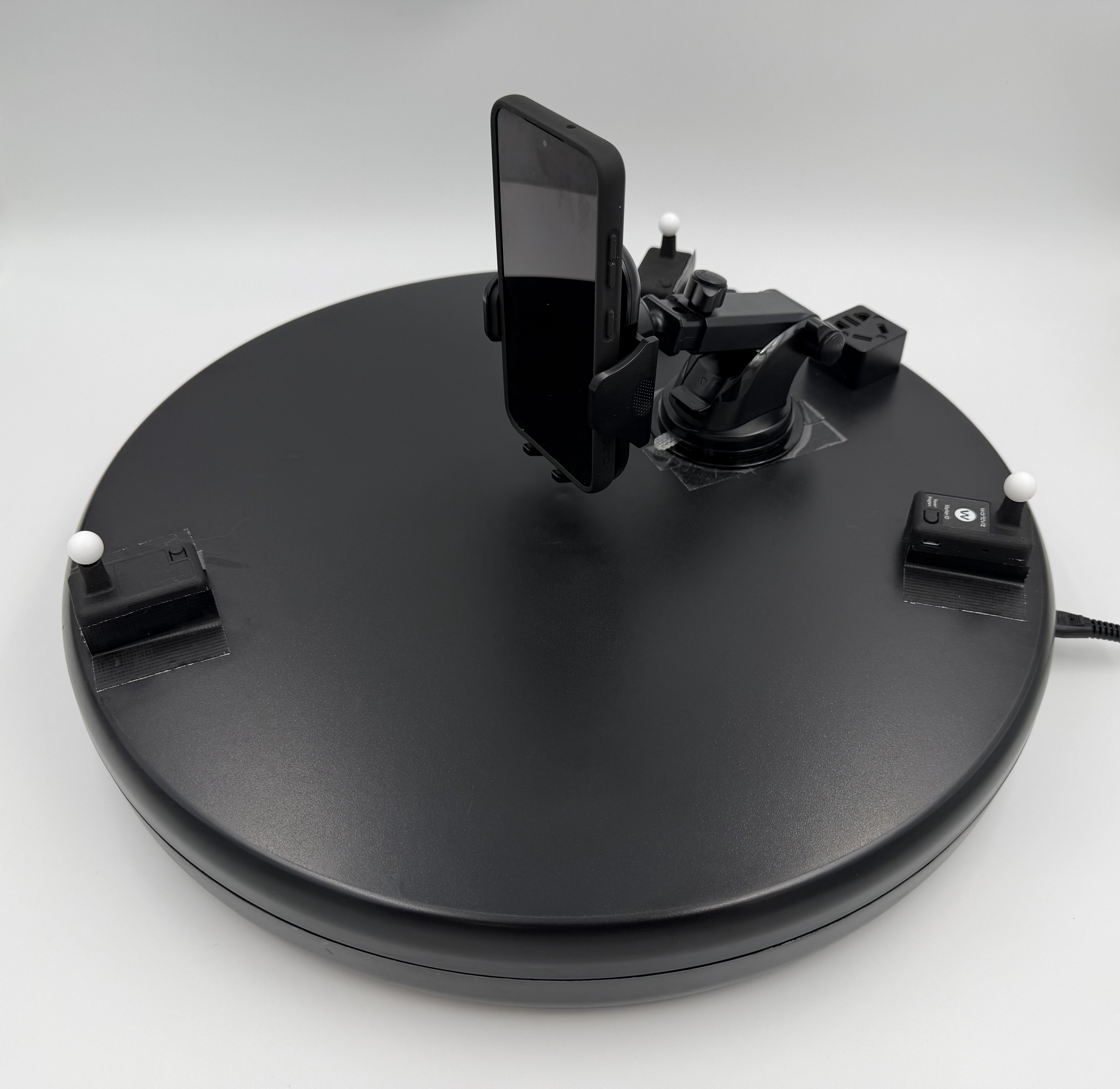}
    \caption{Photo of the rotation table and the \gls{UE} used for CSI measurements.}
    \label{fig:rotation_table_ue}
\end{figure}

During measurements, a \gls{UE} mounted on a rotating table, as shown in \fref{fig:rotation_table_ue}, rotates at a constant angular velocity of approximately $15.8$\,$^\circ$/s. This setup enables \gls{CSI} measurements with varying \gls{UE} orientation at fixed positions. We collected five measurements of the rotating \gls{UE} at four different positions in the measurement area shown in \cite[Fig.~3c]{wiesmayr2025csi}. We recorded the first and the last (fifth) measurement at the same location with a time difference of $18$\,min in order to evaluate \gls{CSI} measurement reproducibility. In the last measurement, we introduced human movement and moved a metallic table in front of \gls{ORU}~6 to investigate the effect of a dynamic \gls{RF} environment.
The CAEZ-5G-ORIENTATION dataset and code will be made publicly available at \url{https://caez.ethz.ch} upon acceptance of this paper.

\begin{table}[tb]
\centering
\caption{Summary of the measured \gls{CAEZ}-5G-ORIENTATION and \gls{CAEZ}-5G-OUTDOOR \gls{CSI} datasets.}
\label{tbl:datasets}
\renewcommand{\arraystretch}{1.1}
\resizebox{0.99\columnwidth}{!}{
\begin{tabular}{@{}lcc@{}}
\toprule
\textbf{\gls{CAEZ}-5G} & \textbf{ORIENTATION} & \textbf{OUTDOOR}\cite{wiesmayr2025csi}\\
\midrule
Duration & $20$\,min  & $1$\,h\,$38$\,min \\
Area & $4$\,m\,$\times4$\,m & $10$\,m\,$\times$\,$10$\,m \\
\# of samples & $132\,981$ & $303\,189$ \\
\gls{UE} type & Samsung Galaxy S23 & Samsung Galaxy S23  \\ 
Vehicle & rotational table & custom robot \\
Position tagged & yes &  yes \\
Rotation tagged & yes & yes \\
PUSCH every & $10$\,ms & $20$\,ms \\
\bottomrule
\end{tabular}
}
\end{table}

\subsection{Orientation Dependence of Measured CSI}

We investigate the orientation dependence of \gls{CSI} by analyzing the subcarrier \gls{CSI} magnitude $\lvert H_\omega [k] \rvert$ of a fixed \gls{ORU} antenna. Here, $\omega$ is the subcarrier index and $k$ is the \gls{OFDM} symbol index; we associate each \gls{OFDM} symbol index with the \gls{UE} orientation from \gls{WorldVizPPT}. For each subcarrier and uplink slot, we average $\lvert H_\omega [k] \rvert$ over three \gls{DMRS} symbols in the same~slot.

\begin{figure}[tb]
    \centering

    \subfloat[\label{fig:csi_dependence_on_rotation_lines}]{
   \includegraphics[width=0.99\columnwidth]{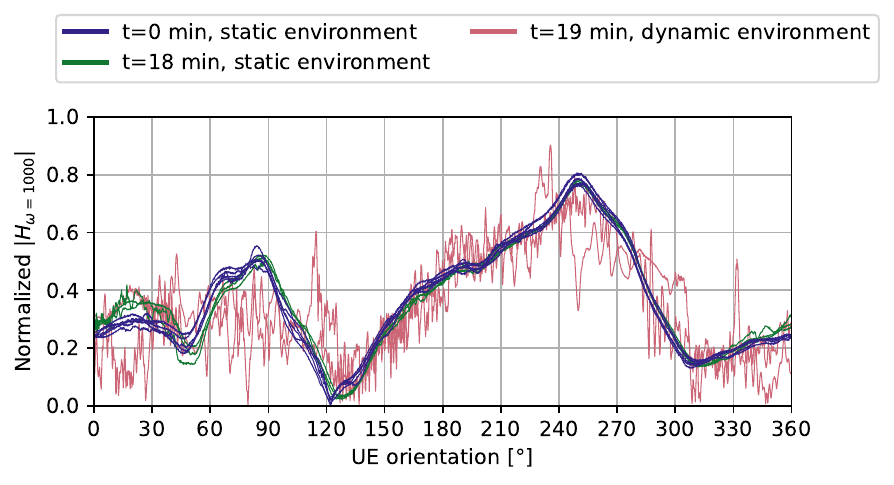}  }

\vspace{-0.3cm}

\subfloat[\label{fig:csi_dependence_on_rotation_waterfall_noise}]{
   \includegraphics[width=0.99\columnwidth]{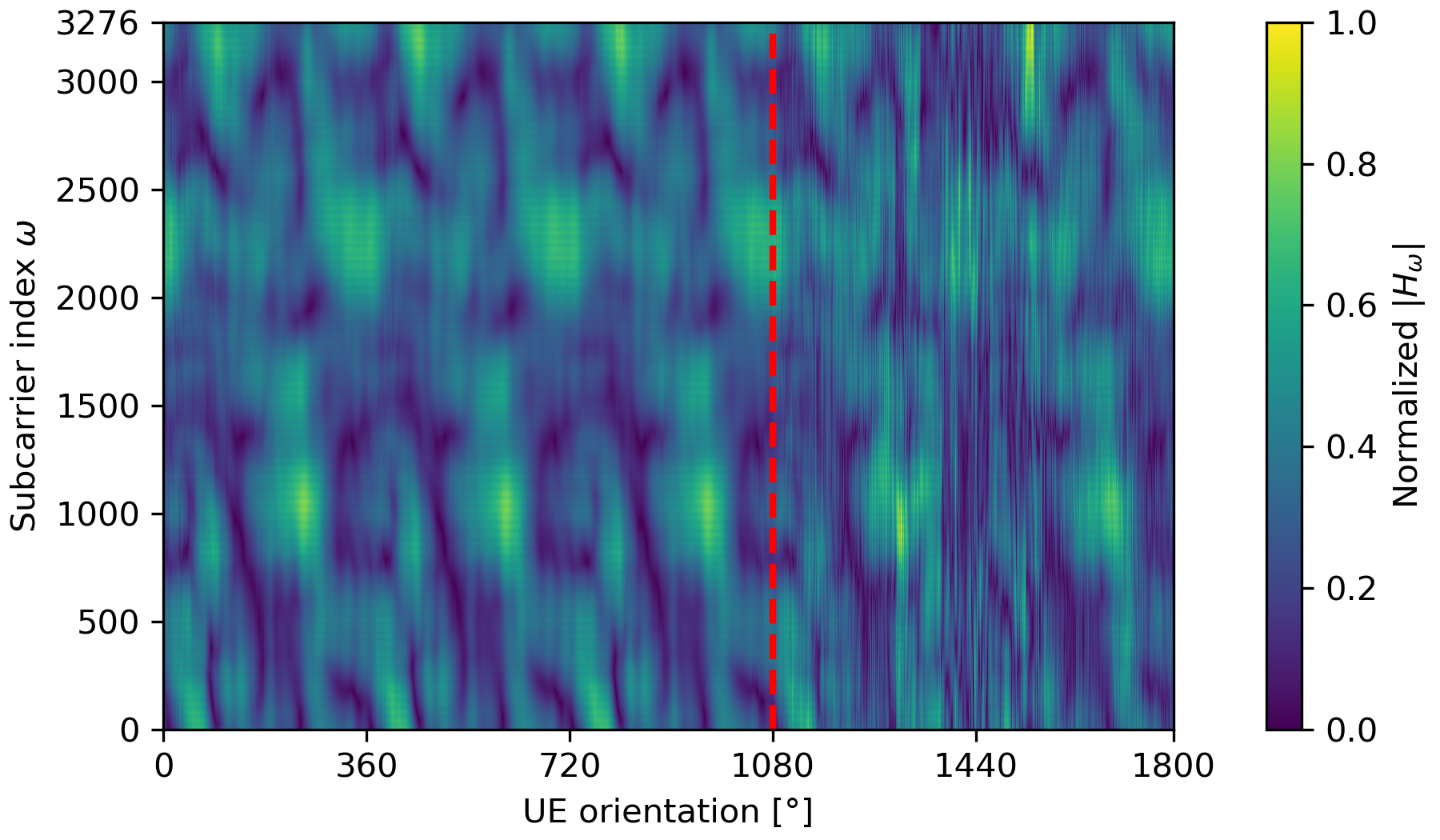} }
    \caption{Dependence of \gls{CSI} on \gls{UE} orientation. (a) \gls{CSI} magnitudes vs. \gls{UE} orientation for O-RU $6$ (cell $51$), antenna $1$, and subcarrier $\omega=1000$, all measured at the same position at different times. (b) \gls{CSI} magnitudes from the fifth measurement for all subcarriers vs. \gls{UE} orientation. The environment was static during the first three rotations (shown on the left of the vertical red line) and dynamic during the final two rotations (shown on the right).}
    \label{fig:csi_dependence_on_rotation}
\end{figure}

\fref{fig:csi_dependence_on_rotation_lines} shows $\lvert H_\omega [k] \rvert$ on subcarrier $\omega=1000$ of the first antenna of \gls{ORU}~6 over all \gls{UE} orientations. The measured \gls{CSI} depends strongly on the \gls{UE} orientation, and this orientation-dependent pattern is consistent over multiple revolutions and remains largely reproducible across repeated measurements. In particular, the measurement recorded $18$\,min later at the same position exhibits a similar \gls{CSI} pattern, which indicates temporal consistency. Human movement and the displacement of a metallic table introduce additional fluctuations, but the dominant orientation-dependent structure remains visible.
\fref{fig:csi_dependence_on_rotation_waterfall_noise} extends the same experiment to all subcarriers, which are visualized on the $y$-axis. The orientation-dependent behavior is visible across all subcarriers. These observations confirm that \gls{UE} orientation information is embedded in measured \gls{CSI}.

\section{Channel Charting for Position and Orientation}
\label{sec:joint_cc}

We now summarize the basics of state-of-the-art \gls{CC} in real-world coordinates \cite{taner2025chartingrealworld} and the triplet loss \cite{ferrand2022tripletbased}, and then detail the steps that extend \gls{CC} to also extract orientation information.

\subsection{Basics of Channel Charting}
\label{sec:basics_of_channel_charting}

\gls{CC} is a self-supervised \gls{ML} method that maps high-dimensional \gls{CSI} to a low-dimensional latent space that resembles the spatial geometry of  \gls{UE} position. Without loss of generality, we focus on two coordinates $(x,z)$ and detail our extension using only the yaw angle $\psi$ as the \gls{UE} orientation.
Our approach could naturally be extended to 3D position and 2D (or 3D) orientation estimation.

\begin{figure}[tp]
    \centering
    \includegraphics[width=0.99\linewidth]{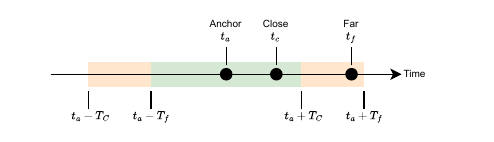}
    \vspace{-0.2cm}
    \caption{Illustration of triplet construction with anchor ($a$), close ($c$), and far~($f$) sample from the time intervals $T_\text{c}$ and $T_\text{f}$; figure inspired by \cite[Fig. 4]{ferrand2022tripletbased}.}
    \label{fig:triplet_selection}
\end{figure}

Triplet-based \gls{CC} \cite{ferrand2022tripletbased} learns a parametric mapping that transforms \gls{CSI} features into a low-dimensional latent space by exploiting structure in the measured training data, specifically, closeness of \gls{UE} position in space and time. Instead of relying on ground-truth position labels for training, this method forms triplets of anchor ($a$), close ($c$), and far ($f$) \gls{CSI} samples, where temporal proximity provides a self-supervised cue on spatial proximity. Therefore, analogous to~\cite{ferrand2022tripletbased}, we form triplets for both position and orientation training based on two time thresholds~$T_\text{c}$ and $T_\text{f}$, which define the close and far intervals, respectively. For a given anchor \gls{CSI} sample, we select close \gls{CSI} samples within the interval $T_\text{c}$ and far \gls{CSI} samples outside the $T_\text{c}$ interval but within the $T_\text{f}$ interval; see \fref{fig:triplet_selection}.

In order to align a channel chart with the physical space, reference \cite{taner2025chartingrealworld} proposes the bilateration loss, which utilizes known \gls{ORU} positions and the relative received signal powers at each \gls{ORU}. The basic idea is to place the low-dimensional embeddings closer to the \gls{ORU} that measures higher receive power. The \gls{CC} function that maps \gls{CSI} features to \gls{UE} positions is typically implemented by a fully-connected \gls{MLP} \cite{ferrand2020dnnbased,taner2025chartingrealworld,wiesmayr2025csi}.

\subsection{Channel Charting for Orientation Estimation}\label{sec:cc_for_orientation_estimation}

In order to extend channel charting to orientation estimation, we build upon the idea of triplet-based \gls{CC}~\cite{ferrand2022tripletbased}. Orientation estimation has one fundamental difference to position estimation in Euclidean space: the orientation angle $\psi$ is $2\pi$-periodic. For triplet-based \gls{CC}, we thus require a measure of angular dissimilarity that accounts for this periodicity. Furthermore, to extract absolute orientation in real-world coordinates, we also require a novel loss function that aligns the orientation estimate $\hat{\psi}$ with the global \gls{UE} orientation.

\subsubsection{Triplet Loss for Orientation Training}

As for conventional triplet-based \gls{CC} \cite{ferrand2022tripletbased}, we assume that \gls{CSI} feature samples close in time have more similar orientations than \gls{CSI} feature samples far in time. Hence, the yaw angle orientation estimate of the anchor sample $\hat{\psi}_\text{a}$ should be closer to the angle of the close sample $\hat{\psi}_\text{c}$ than to the angle of the far sample $\hat{\psi}_\text{f}$ by a margin parameter~$M_\text{O}$, i.e.,
    \begin{align}
    \label{eqn:dist_requirement}
    d(\hat{\psi}_\text{a},\hat{\psi}_\text{c}) + M_\text{O} \leq d(\hat{\psi}_\text{a},\hat{\psi}_\text{f}),
    \end{align}
    for some suitable dissimilarity metric $d$.

In order to design a suitable dissimilarity metric that accounts for the $2\pi$-periodicity of angles, we first convert the angles $\psi_1$ and $\psi_2$ into the unit orientation vectors $\bmo_1 = [\cos(\psi_1), \sin(\psi_1)]^\Tran$ and $\bmo_2 = [\cos(\psi_2), \sin(\psi_2)]^\Tran$. We then compute the following dissimilarity
    \begin{align}
        d(\psi_1,\psi_2) = \textstyle \frac{1}{2} \lVert \bmo_1 - \bmo_2 \rVert^2 = 1 - \langle \bmo_1, \bmo_2 \rangle,
    \end{align}
    where $\langle\cdot,\cdot\rangle$ denotes the inner product. Note that the right-hand side is the cosine dissimilarity between $\bmo_1$ and $\bmo_2$, or, equivalently, in terms of angles: $d(\psi_1,\psi_2) = 1-\cos(\psi_1-\psi_2)$. 

In our implementation, the \gls{CC} function does not output an estimate of the orientation angle $\hat\psi$, but rather an estimate of the \emph{unnormalized} orientation vector $\hat\bmo$, which is not necessarily of unit length.
Our dissimilarity metric is then obtained as follows: For two unnormalized orientation vectors $\hat{\bmo}_1 = [\hat{o}_\text{x,1}, \hat{o}_\text{z,1}]^\Tran$ and $\hat{\bmo}_2 = [\hat{o}_\text{x,2}, \hat{o}_\text{z,2}]^\Tran$, we first normalize these orientation vectors followed by computing the cosine dissimilarity as
    \begin{align}
        \label{eqn:cos_dist}
        d_\text{c}(\hat{\bmo}_1,\hat{\bmo}_2)
        = 
        1 -
        \frac{\langle \hat{\bmo}_1, \hat{\bmo}_2 \rangle}
        {\left\|\hat{\bmo}_1\right\|\left\|\hat{\bmo}_2\right\|}.
    \end{align}
    which is minimal (zero) if the unnormalized orientation vectors~$\hat{\bmo}_1$ and $\hat{\bmo}_2$ are perfectly aligned, i.e., the corresponding orientation angles are equal, and maximal (two) if these orientation vectors are antipodal, i.e., the orientation angles are~$\pi$ apart. We note that unnormalized orientation vectors can be converted into yaw orientation angles using $\hat\psi = \atantwo(\hat{o}_\text{z},\hat{o}_\text{x}).$
    
    Using the dissimilarity from \eqref{eqn:cos_dist}, we define our orientation triplet loss directly on the unnormalized orientation vectors as
    \begin{align}
    \label{eqn:orientation_triplet_loss_cos_dist}
    \mathcal{L}_{\text{tri-}\psi}(\hat{\bmo}_\text{a},\hat{\bmo}_\text{c},\hat{\bmo}_\text{f})
    =
    \left[\,
        d_\text{c}(\hat{\bmo}_\text{a},\hat{\bmo}_\text{c})
        +
        M_\text{O}
        -
        d_\text{c}(\hat{\bmo}_\text{a},\hat{\bmo}_\text{f})\,
    \right]_+,
    \end{align}
    where $[x]_+=\max\{x,0\}$ stands for the \gls{ReLU} activation function.

\subsubsection{Alignment Loss for Absolute Orientation Training}

While the triplet loss in \eqref{eqn:orientation_triplet_loss_cos_dist} enables \gls{CC} for orientation, the resulting angles $\psi$ are not yet aligned to real-world coordinates. To remedy this situation, we use absolute position estimates obtained from \gls{CC} in real-world coordinates \cite{taner2025chartingrealworld}. For the training data only, we assume that the orientation of the \gls{UE} is aligned with its heading, i.e., its motion direction, which itself varies only slowly. As such, the orientations $\bmo_\text{r}$ and $\bmo_\text{n}$ associated with two CSI samples that are close in time, the reference~($r$) and the near~($n$) sample, should be similar.

\begin{figure}[t]
    \centering
    \begin{tikzpicture}[scale=2.0, every node/.style={font=\small}]
        \coordinate (A) at (0,0);
        \coordinate (C) at (2,0.4);

        \draw[thick,-latex,myblue] (A) -- (C) node[midway, above right=0.05cm and -0.05cm] {$\bmc$};

        \filldraw[black] (A) circle (0.02) node[below left] {$\bmp_\text{r}$};
        \filldraw[black] (C) circle (0.02) node[above right] {$\bmp_\text{n}$};

        \draw[thick,-latex,myred] (A) -- ++(0.9,0.25) node[pos=0.65, above] {$\bmo_\text{r}$};
        \draw[thick,-latex,myred] (C) -- ++(0.9,0.05) node[pos=0.65, above] {$\bmo_\text{n}$};

        \coordinate (M) at ($(A)!0.5!(C)$);
        \filldraw[gray] (M) circle (0.03);

        \draw[thick,-latex,myred,dashed] (M) -- ++(0.9,0.15) node[pos=0.8, below] {$\bmo_{\text{avg}}$};

    \end{tikzpicture}
    \vspace{-0.1cm}
    \caption{Visualization of the chord vector $\bmc$ between the two positions~$\bmp_\text{r}$ and~$\bmp_\text{n}$, the orientation vectors $\bmo_\text{r}$ and $\bmo_\text{n}$, and the average orientation $\bmo_{\text{avg}}$. The orientations are closely aligned with the chord vector, which enables us to create an alignment loss that penalizes non-alignment between chord vector~$\bmc$ and average orientation vector $\bmo_{\text{avg}}$.}
    \label{fig:chord_alignment_visualization}
\end{figure}
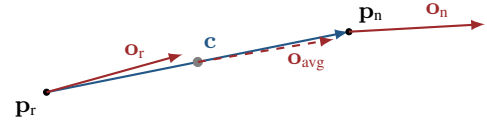

In order to exploit this insight, we define the chord vector between the positions $\bmp_{\text{r}}$ and $\bmp_{\text{n}}$ associated with the reference~($r$) and the near ($n$) sample as follows: $\bmc = \bmp_{\text{n}}-\bmp_{\text{r}}.$
Without loss of generality, we assume that the time associated with the near sample is after the time associated with the reference sample, i.e., $t_\text{n} > t_\text{r}$, so that the chord vector $\bmc$ points in the time-forward direction from the~($r$) to the~($n$) sample.
Thus, the chord vector serves as a proxy for the \gls{UE} heading between the reference~($r$) and the near~($n$) sample. Since we assume that the \gls{UE} heading and \gls{UE} orientation are aligned for the training data, the chord vector $\bmc$ also serves as a proxy for the orientations $\bmo_\text{r}$ and $\bmo_\text{n}$ associated with the reference and the near sample, respectively. When the corresponding positions~$\bmp_{\text{r}}$ and $\bmp_{\text{n}}$ are in real-world coordinates, the chord vector $\bmc$ is globally aligned and is therefore an orientation approximation in real-world coordinates. We illustrate this concept in \fref{fig:chord_alignment_visualization}.

We now use the globally aligned chord vector $\bmc$ to define a loss term that penalizes non-alignment between the orientation estimates and the chord vector. To remain self-supervised, instead of using the ground-truth position labels $\bmp_{\text{r}}$ and $\bmp_{\text{n}}$, we use the position \emph{estimates} $\hat{\bmp}_\text{r}=[\hat{x}_\text{r},\hat{z}_\text{r}]^\Tran$ and $\hat{\bmp}_\text{n}=[\hat{x}_\text{n},\hat{z}_\text{n}]^\Tran$ from the \gls{CC} function in real-world coordinates \cite{taner2025chartingrealworld} to compute an estimate $\hat{\bmc} = \hat{\bmp}_\text{n} - \hat{\bmp}_\text{r}$ of the chord vector $\bmc$. We further compute the (unnormalized) average orientation vector
    \begin{align}
        \hat{\bmo}_\text{avg} = \frac{1}{2}\left(\frac{\hat{\bmo}_\text{r}}{\vecnorm{\hat{\bmo}_\text{r}}} + \frac{\hat{\bmo}_\text{n}}{\vecnorm{\hat{\bmo}_\text{n}}}\right)
    \end{align}
between the orientation \emph{estimates} $\hat{\bmo}_\text{r}=[\hat{o}_\text{x,r},\hat{o}_\text{z,r}]^\Tran$ and $\hat{\bmo}_\text{n}=[\hat{o}_\text{x,n},\hat{o}_\text{z,n}]^\Tran$ from the \gls{CC} function. We then compute the cosine dissimilarity \eqref{eqn:cos_dist} between~$\hat{\bmo}_\text{avg}$ and the estimated chord vector $\hat{\bmc}$, which corresponds to our proposed alignment loss
    \begin{align} \label{eqn:alignment_loss}
        \mathcal{L}_{\text{align-}\psi} = d_\text{c}(\hat{\bmo}_\text{avg}, \hat{\bmc}) = 1 - \frac{ \langle \hat{\bmo}_\text{avg}, \hat{\bmc} \rangle}{\vecnorm{\hat{\bmo}_\text{avg}}\vecnorm{\hat{\bmc}}} .
    \end{align}
This alignment loss is minimal (zero) if the chord vector $\hat{\bmc}$ and the average estimated orientation vector $\hat{\bmo}_\text{avg}$ are perfectly aligned, i.e., the corresponding orientation angles are equal, and maximal (two) if these vectors are antipodal, i.e., the corresponding orientations are $\pi$ apart.

We emphasize that our method remains self-supervised as it relies on the position estimates provided by the proposed channel charting function and not on ground-truth positions.
However, for this concept to work, the channel chart must be sufficiently aligned with real-world coordinates, which we achieve by using the bilateration loss from \cite{taner2025chartingrealworld}.\footnote{The bilateration loss is weakly supervised in a sense that it necessitates knowledge of the \gls{ORU} positions; this is typically known at deployment time.} 

Above, we have described the alignment loss for a reference sample~($r$) and a near sample~($n$). In our implementation, we use the anchor~($a$) sample and the close~($c$) sample from the triplets of the triplet loss \eqref{eqn:orientation_triplet_loss_cos_dist}, in correct temporal order, as the reference~($r$) and the near~($n$) sample, respectively. Furthermore, since errors on short chord vectors have a larger impact on orientation than errors on long chord vectors, we only utilize chord vectors for which $\vecnorm{\hat{\bmc}} > c_{\text{min}}$ for some parameter~$c_{\text{min}}$. 

\subsection{Neural Network Architecture}
\label{sec:neural_network_architecture}

\begin{figure}[tb]
    \centering
    \includegraphics[width=0.99\linewidth]{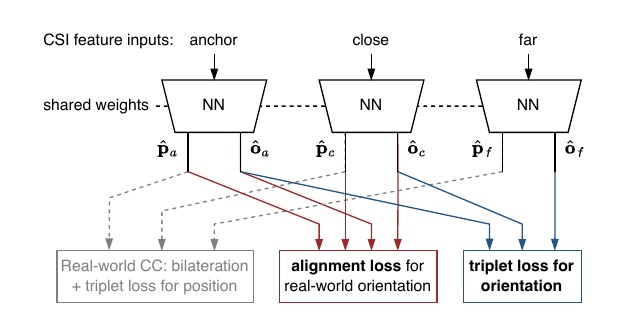}
    \caption{Self-supervised training for joint position and orientation estimation. Our method extends channel charting in real-world coordinates \cite{taner2025chartingrealworld} by an alignment and triplet loss for orientation estimation in real-world coordinates.}
    \label{fig:orientation_losses_combined}
\end{figure}

For joint position and orientation estimation, we adapt the \gls{MLP} architecture from \cite{taner2025chartingrealworld,wiesmayr2025csi}. The neural network consists of seven layers with dimensions $(800,192,96,48,48,48,4)$ and four outputs $[\hat{x},\hat{z},\hat{o}_{\text{x}},\hat{o}_{\text{z}}]^\Tran \in \reals^4$. We use \gls{ReLU} activations for all hidden layers and linear activation for the output layer.
\fref{fig:orientation_losses_combined} illustrates the combination of the \gls{MLP} and the loss function, which consists of the sum of a triplet loss for position~\cite{ferrand2022tripletbased}, a bilateration loss for absolute position in real-world coordinates~\cite{taner2025chartingrealworld}, the proposed triplet loss for orientation~\eqref{eqn:orientation_triplet_loss_cos_dist}, and the proposed alignment loss for orientation~\eqref{eqn:alignment_loss}.

\subsection{Training Method}

To learn the neural network parameters, we use \gls{SGD} to minimize the weighted sum of (i)~the~triplet loss for position from \cite{ferrand2022tripletbased}, (ii) the bilateration and box loss from \cite{taner2025chartingrealworld}, (iii) the triplet loss for orientation \eqref{eqn:orientation_triplet_loss_cos_dist}, and (iv) the real-world alignment loss for orientation \eqref{eqn:alignment_loss}. We scale the individual loss terms by suitable hyperparameters, whose choice we discuss in \fref{sec:training_results_details}.

\section{Results}

\label{sec:results}

\begin{figure*}[tp]
    \centering
    \hfill
    \subfloat[\label{fig:chart_gt}]{
    \includegraphics[width=0.29\textwidth]{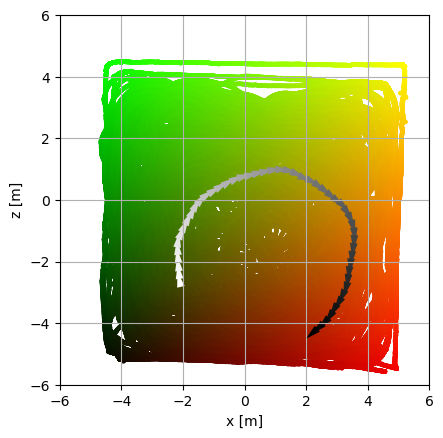}    }
    \hfill
    \subfloat[\label{fig:chart_supervised_estimation}]{
 \includegraphics[width=0.29\textwidth]{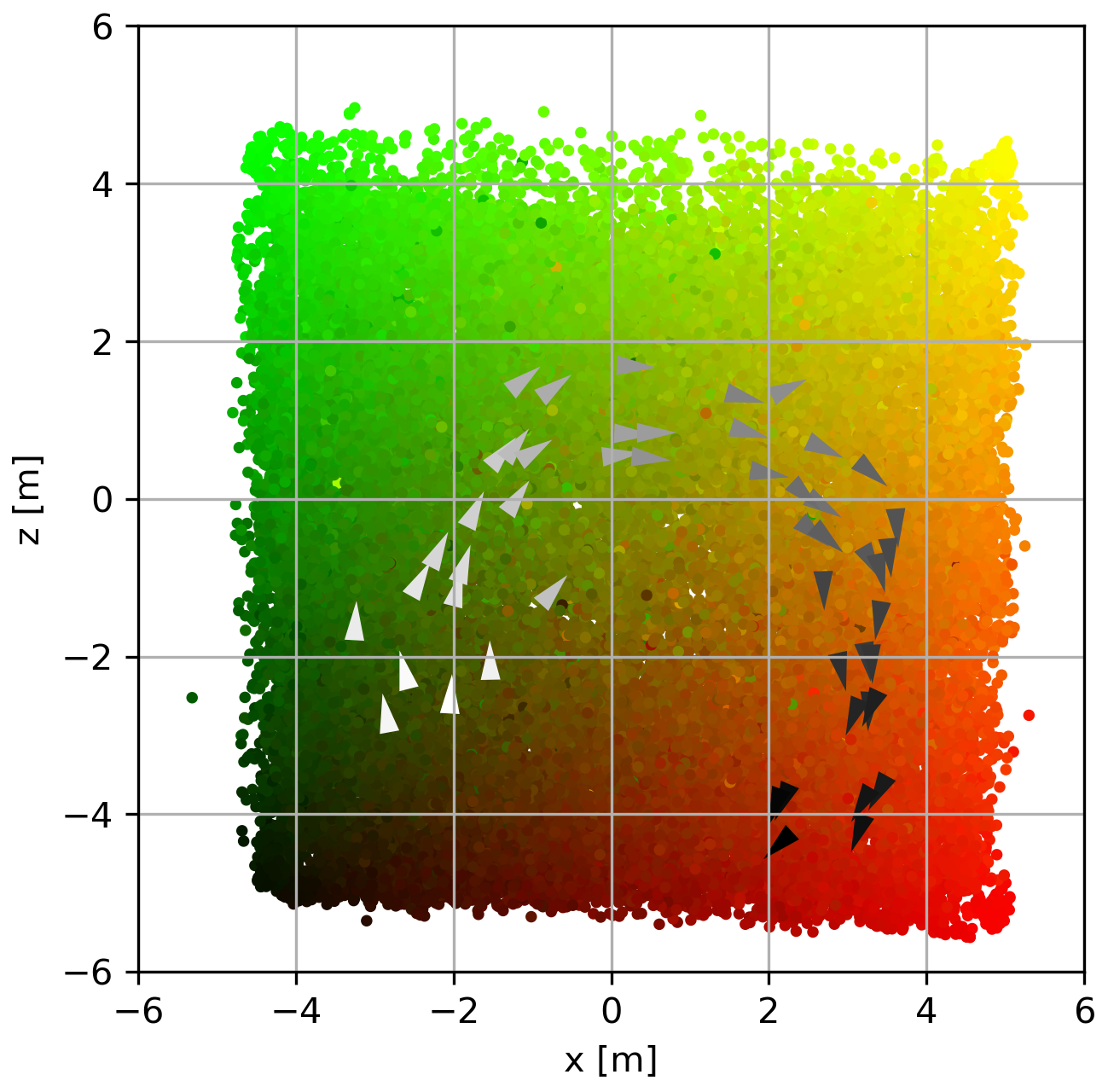}    }
 \hfill
  \subfloat[\label{fig:chart_unsupervised_estimation}]{
        \includegraphics[width=0.29\textwidth]{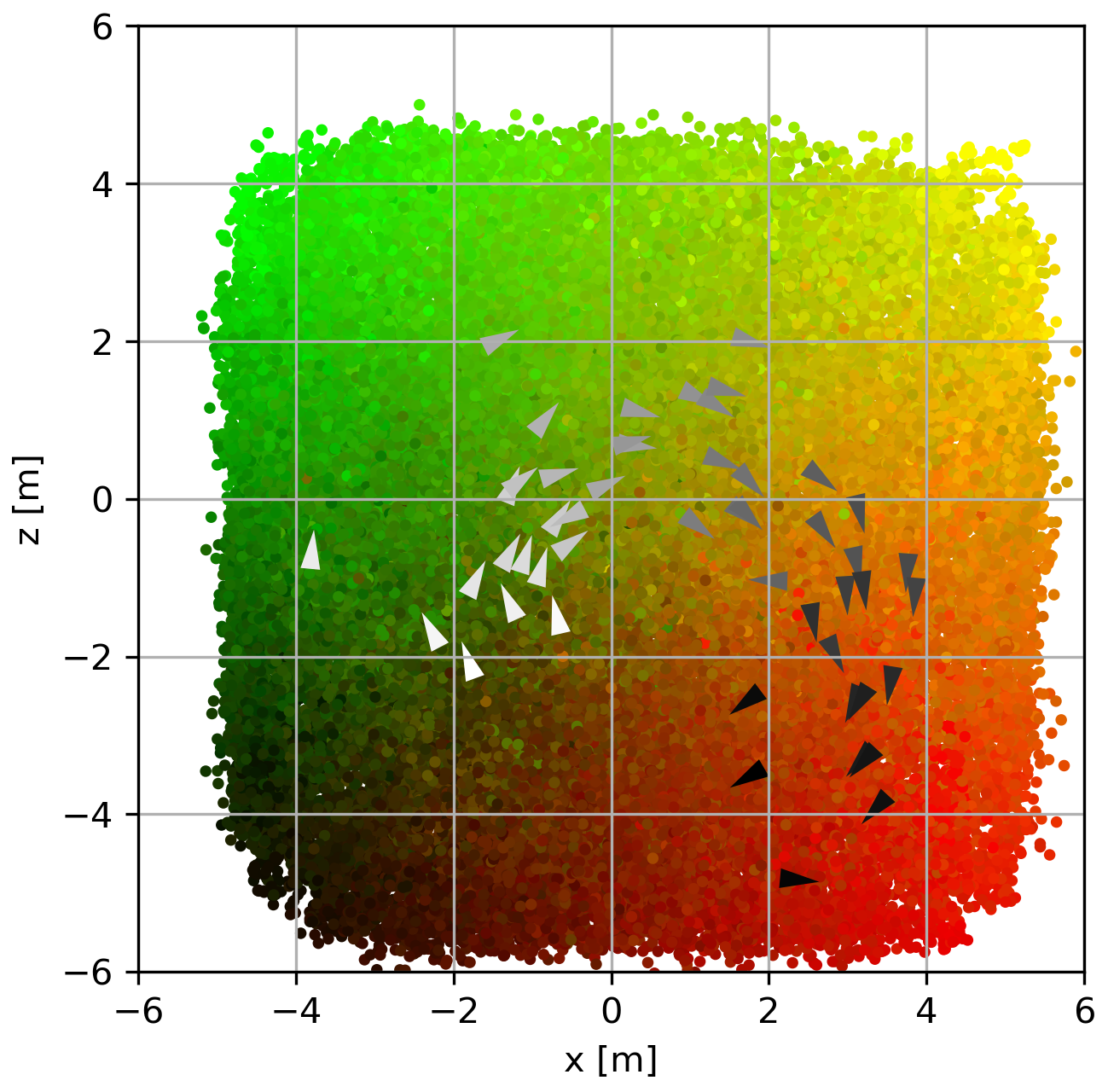} }
    \hfill
    \hspace{0.2cm}

    \caption{Ground truth-positions and orientations (a), results of the neural positioning and orientation estimation baseline (b), and results of the proposed \gls{CC} for position and orientation method~(c). The grayscale arrows indicate the positions and orientations of every $20$th sample of the last $1000$ test samples.}
    \label{fig:charts}
\end{figure*}

We now evaluate our method using measured \gls{CSI} from the CAEZ-5G-OUTDOOR real-world 5G NR dataset \cite{wiesmayr2025csi}.

\subsection{Evaluation Methods}

\subsubsection{Training and Testing Dataset}

We use the publicly available CAEZ-5G-OUTDOOR dataset, which was measured at ETH Zurich in an outdoor courtyard environment using a Samsung Galaxy S23 \gls{UE} carried by a remote-controlled robot vehicle. We summarize the key features of the dataset in \fref{tbl:datasets} and refer to~\cite{wiesmayr2025csi} for the details. During \gls{CSI} acquisition, the robot continuously moved forward on smooth trajectories and the \gls{UE} heading was aligned with the direction of motion.
Thus, we can directly mine triplets from the \gls{CSI} time series, as described in \fref{sec:cc_for_orientation_estimation}. As \gls{CSI} features, we use down-sampled \gls{OFDM}-domain absolute values, as described in \cite[Sec.~IV-B2]{wiesmayr2025csi}. We exclude the last $10^4$ samples from the entire dataset and refer to them as the ``test dataset.'' We then partition the remaining data into a training and a validation dataset, applying a random split of 80\,\% to 20\,\%, respectively.

\subsubsection{Performance Metrics}
\label{sec:performance_metrics}

To evaluate position estimation quality, we measure \gls{TW}, \gls{CT}, and \gls{KS}~\cite{studer2018channelcharting}. We also report the mean, median, and 95th percentile of the per-sample distance error
    \begin{align}
        e_{\text{pos},i} = \vecnorm{ \hat{\bmp}_i - \bmp_i }
    \end{align}
between estimated position $\hat{\bmp}_i$ and ground-truth position~$\bmp_i$, where $i$ is the sample index. Ground-truth position information is only used for performance evaluation---not for training.

To evaluate orientation estimation quality, we use the mean, median, and 95th percentile over all per-sample absolute angle errors. Concretely, we first estimate the angle $\hat{\psi}_i = \atantwo(\hat{o}_{\text{z},i}, \hat{o}_{\text{x},i})$ and then report the minimum angle error
    \begin{align}
        \label{eq:absolute_error_orientation}
        e_{\psi,i} =\min_{n\in\mathbb{Z}} \big|\hat{\psi}_i-\psi_i-2\pi n\big|,
    \end{align}
which we finally convert into degrees. This performance metric allows the following interpretation: (i)~identical orientations yield an error of $0^\circ$, (ii)~uniformly random orientations yield an expected error of $90^\circ$, and (iii)~antipodal orientations yield an error of $180^\circ$. We evaluate all metrics, with respect to both position and orientation, from single-shot estimates on both test datasets, i.e., the validation dataset samples from random partitioning and the test dataset from the last $10^4$ samples.
                
\subsubsection{Baseline Methods}

As a baseline for positioning only, we use \emph{supervised} neural positioning and \emph{self-supervised} \gls{CC} \emph{without orientation estimation}, as implemented in \cite{wiesmayr2025csi}. For neural positioning, we use an \gls{MLP} that directly outputs position estimates, which we train using a \acrfull{MSE} loss and ground-truth position labels. As a baseline for joint position and orientation estimation, we also use \emph{supervised} neural positioning \emph{with orientation estimation}. To this end, we train the \gls{NN} proposed in \fref{sec:neural_network_architecture} with ground-truth position and orientation labels.
From the ground-truth orientation $\psi_i$, we compute the orientation $[\cos(\psi_i), \sin(\psi_i)]^\Tran$ and apply the cosine dissimilarity loss from~\eqref{eqn:cos_dist}.

\subsubsection{Training Methods}\label{sec:training_results_details}

We train the \gls{MLP} described in \fref{sec:neural_network_architecture} using PyTorch \cite{pytorch}, the Adam optimizer \cite{kingma2017adammethodstochasticoptimization}, and a StepLR learning-rate scheduler. We train the network for 200 epochs with a batch size of 2048 anchors, where for each anchor we mine two triplets, i.e., two close and two far samples. We obtain a well-performing set of hyperparameters using the Optuna optimization framework \cite{optuna_2019} together with the Optuna AutoSampler \cite{ozaki2025optunahub}. The hyperparameter optimization involves the weights of all loss terms, the loss margins, the close and far time thresholds used for triplet mining, the learning rate, and the scheduler step size. The optimization objective comprises \gls{TW}, \gls{CT}, \gls{KS}, mean absolute errors of position and orientation, and the cosine distance between predicted and ground-truth orientations.
Based on these objectives, Optuna constructs a Pareto-optimal front using non-dominated sorting. We use the validation dataset for hyperparameter optimization, while the additional test data (the last $10^4$ samples) is used for performance evaluation only.

\begin{table*}[tp]
\centering
\caption{Results of \gls{CC} for position and orientation compared to baseline methods ($\shortuparrow$ indicates large is good; $\shortdownarrow$ indicates small is good)}
\label{tbl:results_overview}
\renewcommand{\arraystretch}{1.1}
\resizebox{0.99\textwidth}{!}{
\begin{tabular}{@{}lccccccccc@{}}\toprule
& \multirow{2}{*}{TW\,$\shortuparrow$} & \multirow{2}{*}{CT\,$\shortuparrow$} & \multirow{2}{*}{KS\,$\shortdownarrow$} & \multicolumn{3}{c}{Absolute error of  $\hat{\bmp}$ [cm] $\shortdownarrow$} &  \multicolumn{3}{c}{Min. angle error $e_{\psi}$ [\textdegree] $\shortdownarrow$}  \\
& & & & Mean & Median & 95th\,\% &  Mean & Median & 95th\,\% \\
\midrule
\textbf{Evaluated on validation dataset} \\
Baseline neural positioning & 0.992 & 0.992 & 0.077 & 30 & 23 & 79 &   -- & -- & -- \\
Baseline channel charting & 0.982 & 0.974 & 0.167 & 73 & 64 & 158 &  -- & -- & -- \\
Baseline joint neural pos. with orientation & 0.992 & 0.992 & 0.080 & 32 & 25 & 79 &  14 & 7.3 & 42 \\
\gls{CC} for position and orientation (ours) & 0.975 & 0.974 & 0.160 & 75 & 68 & 156 &  20 & 11 & 71 \\
\midrule
\textbf{Evaluated on test dataset (last $\bf 10^4$ samples)} \\
Baseline joint neural pos. with orientation & 0.971 & 0.973 & 0.148 & 50 & 39 & 127 & 20 & 8.9 & 125 \\
\gls{CC} for position and orientation (ours) & 0.960 & 0.957 & 0.207 & 82 & 74 & 168 &  22 & 11 & 108 \\
\bottomrule
\end{tabular}}
\end{table*}

\subsection{Position Estimation Results}

We summarize the results of our experiments in \fref{tbl:results_overview}. We observe that joint position and orientation estimation achieves positioning performance similar to that of the position-only counterparts. Compared to the results reported in \cite{wiesmayr2025csi}, we see slightly worse \gls{TW} but similar \gls{CT} and slightly better \gls{KS}, while the absolute error of the estimated positions increases only slightly. \fref{fig:charts} compares the ground-truth positions in \fref{fig:chart_gt} with the estimates from the joint neural positioning and orientation estimation baseline in \fref{fig:chart_supervised_estimation}, and with the estimates from the proposed \gls{CC} for position and orientation method in \fref{fig:chart_unsupervised_estimation}. We color-code the samples from the validation dataset by a green-to-red color gradient. Furthermore, we visualize a trajectory of every $20$th of the last $1\,000$ test samples by a white-to-black color gradient, which captures both \gls{UE} position and orientation. \fref{fig:chart_unsupervised_estimation} exhibits higher estimation noise than the supervised baseline shown in \fref{fig:chart_supervised_estimation}.

\subsection{Orientation Estimation Results}

\begin{figure}[tp]
    \centering
     \includegraphics[width=0.95\columnwidth]{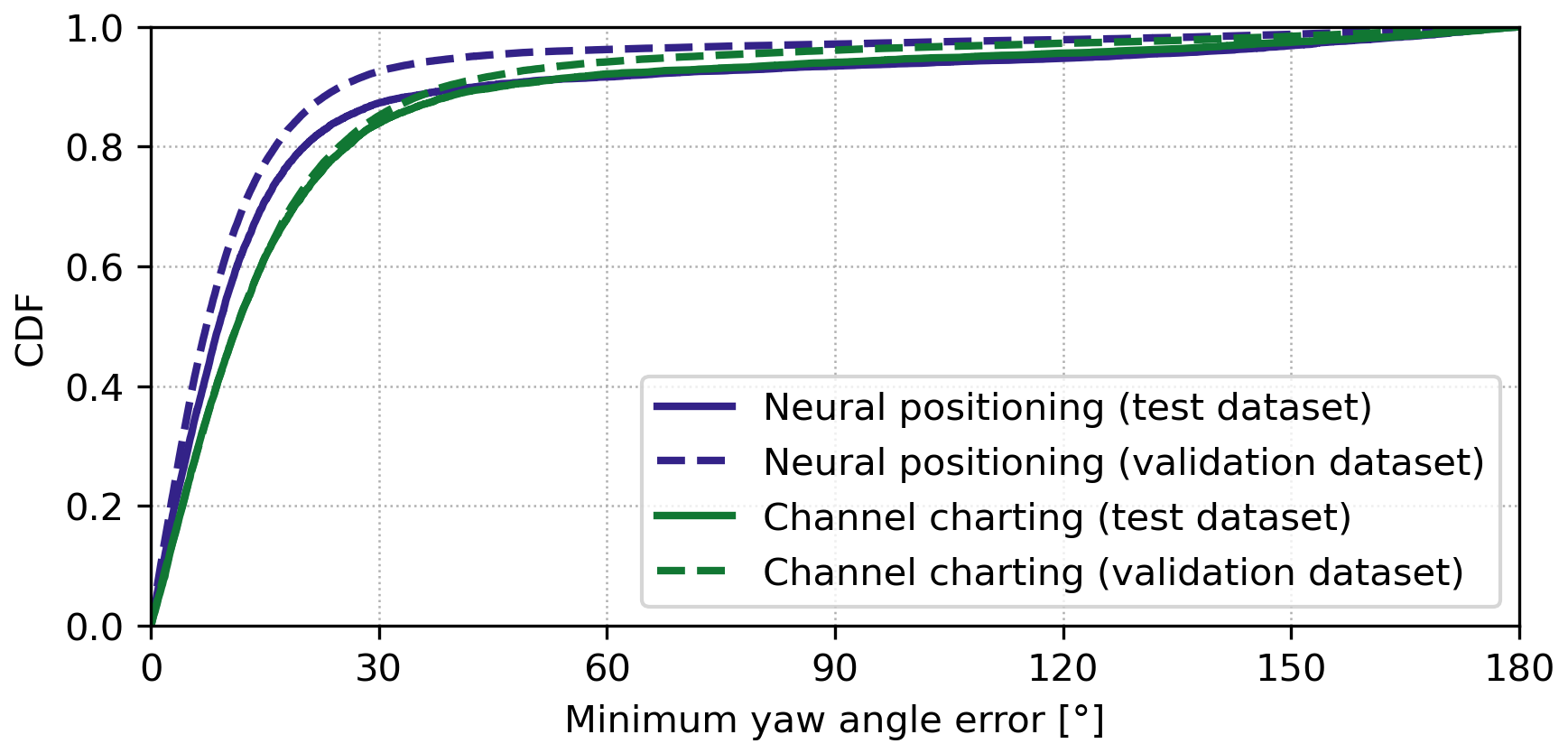}
    \caption{\Acrfull{CDF} of the minimum yaw angle error for the joint neural positioning and orientation estimation baseline, as well as for the proposed \gls{CC}-based method. The \glspl{CDF} were computed for the test and validation datasets with respect to the ground-truth yaw angles.}
    \label{fig:angle_vs_time}
\end{figure}

We summarize the results of our experiments in \fref{tbl:results_overview}. For yaw estimation on the test dataset, orientation-aware \gls{CC} achieves orientation estimation performance comparable to that of the joint neural positioning and orientation estimation pipeline, with a difference of only~$2$\textdegree\ in mean minimum angle error. \fref{fig:charts} compares the ground-truth orientations in \fref{fig:chart_gt} with the estimates from the joint neural positioning and orientation estimation baseline in \fref{fig:chart_supervised_estimation}, and with the estimates from the proposed \gls{CC} for position and orientation method in \fref{fig:chart_unsupervised_estimation}. Furthermore, \fref{fig:angle_vs_time} shows the \gls{CDF} of the minimum angle error in \eqref{eq:absolute_error_orientation} with respect to the ground-truth yaw angle, both for neural positioning and orientation estimation and for \gls{CC} for position and orientation, evaluated on the validation and test datasets.
\fref{fig:angle_vs_time} demonstrates that the proposed \gls{CC} method for position and orientation closely approaches the accuracy of supervised joint neural positioning and orientation estimation while being trained in a self-supervised fashion---without the use of any ground-truth position or orientation labels.

\section{Conclusions}
\label{sec:conclusions}

We have extended state-of-the-art \gls{CC} in real-world coordinates~\cite{taner2025chartingrealworld} to joint single-shot estimation of \gls{UE} position and orientation. Our training method aligns the orientation to real-world coordinates in a self-supervised manner by using the position estimates in real-world coordinates obtained from the bilateration loss. Our experiments with real-world \gls{CSI} measurements have revealed that the orientation estimation performance closely approaches that of a supervised neural positioning baseline. Furthermore, the proposed \gls{CC} for position and orientation method comes at only a negligible degradation in positioning performance. These results demonstrate that resolving the orientation dependence of measured \gls{CSI} further improves the expressivity of \gls{CC}, with the potential to assist and improve beamfinding, precoding, beam- and cell-assignment, \gls{UE} trajectory prediction, and many more tasks.

\balance
\bibliographystyle{IEEEtran}

\balance

\end{document}